\documentclass[12pt]{article}
\usepackage{epsfig}
\begin{document}
%%%%%%%%%%%%%%%%%%%%%%%%%%%%%%%%%%%%%%%%%%%%%%%%%%%%%%%%%%%%%%%%%%%%%%%%%
\title{ Near-threshold $\eta^\prime$ meson production in ${\pi^-}A$ reactions}
\author{E. Ya. Paryev$^{1,2}$\\
{\it $^1$Institute for Nuclear Research, Russian Academy of Sciences,}\\
{\it Moscow 117312, Russia}\\
{\it $^2$Institute for Theoretical and Experimental Physics,}\\
{\it Moscow 117218, Russia}}
%%==============================================================
%%==============================================================

\renewcommand{\today}{}
\maketitle

\begin{abstract}
We study the inclusive near-threshold production of $\eta^\prime$ mesons
in $\pi^-$ meson-nucleus reactions on the basis of the first-collision model relying on the
nuclear spectral function and including incoherent processes of $\eta^\prime$ production in
$\pi^-$ meson--proton collisions. The model accounts for the absorption of initial $\pi^-$ and final
$\eta^\prime$ mesons, the binding of intranuclear protons and their Fermi motion, as well as
the effect of the scalar $\eta^\prime$--nucleus potential (or the in-medium shift of the $\eta^\prime$
meson mass) on these processes. We calculate the differential and total cross
sections for $\eta^\prime$ production off carbon and tungsten nuclei at laboratory angles of
0$^{\circ}$--10$^{\circ}$ and 10$^{\circ}$--45$^{\circ}$ by $\pi^-$ mesons with momentum of 1.7 GeV/c,
which is close to the threshold momentum for $\eta^\prime$ creation on a free target proton at rest.
We show that the calculated $\eta^\prime$ production cross sections are larger than those, studied earlier
in the $\eta^\prime$ photoproduction reactions, by about two orders of magnitude.
We also demonstrate that the high absolute values of the present differential
and total $\eta^\prime$ production cross sections in the
momentum ranges of 0.2--0.3 GeV/c and 0.1--0.6 GeV/c, respectively, possess a high sensitivity
to changes in the in-medium shift of the $\eta^\prime$ mass. This offers the possibility of evaluating
the above shift at these momenta from the respective data,
which could be taken in future experiments at the GSI pion beam facility or at J-PARC.
\end{abstract}

\newpage

\section*{1 Introduction}

\hspace{0.5cm} Studies of the $\eta^\prime$ meson mass and its interaction in finite nuclei have received
considerable interest in recent years due to the hope to extract valuable information on the partial
restoration of chiral symmetry at finite density, on gluon dynamics in low-energy QCD, and on the possible
formation of $\eta^\prime$ bound states inside the nuclei (see, for example, Refs.~[1, 2]).
Recent measurements of inclusive $\eta^\prime$ photoproduction from carbon [3] and niobium [4] nuclear targets
by the CBELSA/TAPS Collaboration in Bonn gave for the real part $V_0$ of the $\eta^\prime$--nucleus potential
at saturation density $\rho_0$ (which is equal to the $\eta^\prime$ meson mass shift in the medium) the combined
value of $V_0=-(39\pm7_{stat}\pm15_{syst})$ MeV [1, 4] for average momentum of the produced $\eta^\prime$ of
$\approx$ 1.1 GeV/c. A new determination of the $\eta^\prime$--nucleus potential $V_0$ with lower average
$\eta^\prime$ momentum of 600 MeV/c was performed also by the CBELSA/TAPS Collaboration [5] by measuring the
production of $\eta^\prime$ mesons in coincidence with forward-going protons in photon-induced reactions on
$^{12}$C for initial photon energies of 1.3--2.6 GeV at the electron accelerator ELSA with the result
$V_0=-(44\pm16_{stat}\pm15_{syst})$ MeV, consistent with the above value for the potential $V_0$, deduced
from inclusive measurements for average $\eta^\prime$ momentum of $\approx$ 1.1 GeV/c. The imaginary part
$W_0$ of the $\eta^\prime$--nucleus optical potential was determined in [6] to be
$W_0=-(13\pm3_{stat}\pm3_{syst})$ MeV near-threshold from an extrapolation of the
transparency ratio measurements on Nb at various $\eta^\prime$ momenta well above threshold
to near-threshold momenta. These numbers, with a
small modulus of the imaginary $\eta^\prime$ meson potential compared to the modulus of its real potential
suggest that $\eta^\prime$--nucleus bound states may exist, provided that the above relation between the
imaginary and real potentials persists at low $\eta^\prime$ momenta as well. Thus, the knowledge of the real
part of the $\eta^\prime$--nucleus potential for meson momenta small compared to its mass is crucial for
understanding the feasibility of forthcoming experiments aiming at the search for $\eta^\prime$ mesic bound states.
The search of such states in the $^{12}$C($p,d$) reaction near the $\eta^\prime$ emission threshold has been
recently undertaken at FRS/GSI [7]. No distinct structures, associated with the formation of $\eta^\prime$
bound states, were observed. In the near future, the search of $\eta^\prime$ mesic nuclei is also planned at
GSI/FAIR [7] as well as in photonuclear reactions at ELSA [8].

    In the context of aforesaid, it is interesting to consider another reactions such as
${\pi^-}A \to {\eta^\prime}X$ for clarifying the possibility of constraining in them the $\eta^\prime$ meson
real potential at low momenta. In this respect, the main purpose of the present work is to give the predictions
for the differential and total cross sections for $\eta^\prime$ production in ${\pi^-}^{12}C \to {\eta^\prime}X$
and ${\pi^-}^{184}W \to {\eta^\prime}X$ reactions at 1.7 GeV/c beam momentum.
These nuclear targets and this initial momentum were employed in recent measurements [9]
of $\pi^-$ meson-induced
$\phi$ meson production at the GSI pion beam facility [10] and, therefore, can be used in studying the
${\pi^-}A \to {\eta^\prime}X$ reactions again at GSI or at J-PARC
\footnote{$^)$Especially in view of the fact that there are proposals at J-PARC which plan to use
the ($\pi^-,n$) reaction on nuclear targets to look for $\eta$ [11] and $\omega$ [12] bound states in
nuclei.}$^)$
.
The calculations are based on a first-collision model, developed in [13] for the analysis of the inclusive
$\phi$ meson production data [9] and expanded to take into account different scenarios for the
$\eta^\prime$ meson in-medium mass shift.
When the respective data from the future experiments will become available, the results
can be used as an important tool for determining this shift at low meson momenta.

\section*{2 Framework: a direct  $\eta^\prime$ meson knock out process}

\hspace{0.5cm} The direct production of $\eta^\prime$ mesons
in $\pi^-A$ ($A=^{12}$C and $^{184}$W) collisions at incident momentum of 1.7 GeV/c
can occur in the following $\pi^-p$ elementary process with zero pions in the final state
\footnote{$^)$The free threshold momentum for this process is 1.43 GeV/c.
We can ignore the processes $\pi^-N \to {\eta^\prime}N{\pi}$ with one pion in the final state
at the incident momentum of interest, because this momentum is less than their production
threshold momenta in free $\pi^-N$ interactions. Thus, for example, the threshold momentum of the
channel $\pi^-p \to {\eta^\prime}n{\pi^0}$ is 1.72 GeV/c. This momentum is larger than the incident pion
momentum of 1.7 GeV/c. Consequently, the processes $\pi^-N \to {\eta^\prime}N{\pi}$
are energetically suppressed. Moreover, accounting for the results of the study [13] of pion-induced
$\phi$ meson production on $^{12}$C and $^{184}$W target nuclei at pion beam momentum of 1.7 GeV/c,
we neglect in the present work by analogy with [13] the secondary pion--nucleon ${\pi}N \to {\eta^\prime}N$
production processes.}$^)$
%formula(1)
\begin{equation}
\pi^-+p \to \eta^\prime+n.
\end{equation}
Following Ref. [14], we simplify the subsequent calculations via accounting for
in-medium modification of $\eta^\prime$ mesons, involved in process (1),
in terms of their average in-medium mass
$<m^*_{\eta^\prime}>$, which is defined as:
%formula(2)
\begin{equation}
<m^*_{\eta^\prime}>=m_{\eta^\prime}+V_0\frac{<{\rho_N}>}{{\rho_0}}.
\end{equation}
Here, $m_{\eta^\prime}$ is the $\eta^\prime$ rest mass in free space,
$V_0$ is the $\eta^\prime$ effective scalar
nuclear potential (or its in-medium mass shift) at normal nuclear matter
density ${\rho_0}$, and $<{\rho_N}>$ is the average nucleon density.
For target nuclei $^{12}$C and $^{184}$W,  the ratio $<{\rho_N}>/{\rho_0}$, was chosen as
0.55 and 0.76, respectively, in our ensuing study.
The total energy $E^\prime_{\eta^\prime}$ of the $\eta^\prime$ meson in nuclear matter is
expressed via its average effective mass $<m^*_{\eta^\prime}>$ and its in-medium momentum
${\bf p}^{\prime}_{\eta^\prime}$ by the equation [14]:
%formula(3)
\begin{equation}
E^\prime_{\eta^\prime}=\sqrt{({\bf p}^{\prime}_{\eta^\prime})^2+(<m^*_{\eta^\prime}>)^2}.
\end{equation}
The momentum ${\bf p}^{\prime}_{\eta^\prime}$ is related to the vacuum $\eta^\prime$
momentum ${\bf p}_{\eta^\prime}$ as follows [14]:
%formula(4)
\begin{equation}
E^\prime_{\eta^\prime}=\sqrt{({\bf p}^{\prime}_{\eta^\prime})^2+(<m^*_{\eta^\prime}>)^2}=
\sqrt{{\bf p}^2_{\eta^\prime}+m^2_{\eta^\prime}}=E_{\eta^\prime},
\end{equation}
where $E_{\eta^\prime}$ is the $\eta^\prime$ total energy in vacuum.

 As indicated above, the potential depth $V_0$ of about -40 MeV was deduced in semi-exclusive [5]
and inclusive [3, 4] $\eta^\prime$ photoproduction experiments for average $\eta^\prime$ momenta
of 0.6 and 1.1 GeV/c, respectively. Therefore, in this work it is natural to use the value
$V_{0}=-40$ MeV for the $\eta^\prime$ mass shift $V_0$ at density ${\rho_0}$ for all studied
$\eta^\prime$ momenta. In order to see the sensitivity of the
$\eta^\prime$ production cross sections from the process (1) to the $\eta^\prime$ mass shift
$V_{0}$, we will also both ignore it in our calculations and will adopt the value
$V_{0}=-80$ MeV, predicted by the linear sigma model [15]. It should be pointed out that we restrict
ourselves to the above values for the potential depth $V_0$ due to the fact that strongly attractive
potentials $V_0 \le -100$ MeV were excluded (for a relatively shallow imaginary potentials) by the
experiment of Tanaka et al. [7] (see, also [1]). For the sake of numerical simplicity, in the present
work in evaluating the $\eta^\prime$ production cross sections we will disregard the modification
of the mass of the neutrons, produced together with the $\eta^\prime$ mesons in reaction channel (1),
in the nuclear mean field [16].

  Because the $\eta^\prime$--nucleon elastic cross section is expected to be small [17],
we will neglect quasielastic ${\eta^\prime}N$ rescatterings in the present study.
Then, taking into consideration
the distortion of the incident pion in nuclear matter and describing the $\eta^\prime$ meson final-state
absorption by the effective in-medium momentum-independent ${\eta^\prime}N$ inelastic cross section $\sigma_{{\eta^\prime}N}$
\footnote{$^)$For this cross section we adopt the value $\sigma_{{\eta^\prime}N}=13$ mb motivated by
the results from the $\eta^\prime$ photoproduction CBELSA/TAPS experiment [6].}$^)$
as well as using the results presented in Ref.~[13], we represent the inclusive differential
cross section for the production of ${\eta^\prime}$ mesons with vacuum momentum ${\bf p}_{\eta^\prime}$
on nuclei in the direct process (1) as follows:
%formula(5)
\begin{equation}
\frac{d\sigma_{{\pi^-}A\to {{\eta^\prime}}X}^{({\rm prim})}
({\bf p}_{\pi^-},{\bf p}_{\eta^\prime})}
{d{\bf p}_{{\eta^\prime}}}=I_{V}[A,\theta_{{\eta^\prime}}]
\left(\frac{Z}{A}\right)\left<\frac{d\sigma_{{\pi^-}p\to {{\eta^\prime}}n}({\bf p}_{\pi^-},
{\bf p}^{\prime}_{{\eta^\prime}})}{d{\bf p}^{\prime}_{{\eta^\prime}}}\right>_A
\frac{d{\bf p}^{\prime}_{{{\eta^\prime}}}}{d{\bf p}_{{{\eta^\prime}}}},
\end{equation}
where
%formula(6)
\begin{equation}
I_{V}[A,\theta_{{\eta^\prime}}]=A\int\limits_{0}^{R}r_{\bot}dr_{\bot}
\int\limits_{-\sqrt{R^2-r_{\bot}^2}}^{\sqrt{R^2-r_{\bot}^2}}dz
\rho(\sqrt{r_{\bot}^2+z^2})
\exp{\left[-\sigma_{{\pi^-}N}^{\rm tot}A\int\limits_{-\sqrt{R^2-r_{\bot}^2}}^{z}
\rho(\sqrt{r_{\bot}^2+x^2})dx\right]}
\end{equation}
$$
\times
\int\limits_{0}^{2\pi}d{\varphi}\exp{\left[-\sigma_{{\eta^\prime}N}A
\int\limits_{0}^{l(\theta_{{\eta^\prime}},\varphi)}
\rho(\sqrt{x^2+2a(\theta_{{\eta^\prime}},\varphi)x+b+R^2})dx\right]},
$$
%formula(7)
\begin{equation}
a(\theta_{{\eta^\prime}},\varphi)=z\cos{\theta_{{\eta^\prime}}}+
r_{\bot}\sin{\theta_{{\eta^\prime}}}\cos{\varphi},\\\
b=r_{\bot}^2+z^2-R^2,
\end{equation}
%formula(8)
\begin{equation}
l(\theta_{{\eta^\prime}},\varphi)=\sqrt{a^2(\theta_{{\eta^\prime}},\varphi)-b}-
a(\theta_{{\eta^\prime}},\varphi),
\end{equation}
%formula(9)
\begin{equation}
\left<\frac{d\sigma_{{\pi^-}p\to {\eta^\prime}n}({\bf p}_{\pi^-},{\bf p}_{\eta^\prime}^{'})}
{d{\bf p}_{\eta^\prime}^{'}}\right>_A=
\int\int
P_A({\bf p}_t,E)d{\bf p}_tdE
\end{equation}
$$
\times
\left\{\frac{d\sigma_{{\pi^-}p\to {\eta^\prime}n}[\sqrt{s},<m^*_{{\eta^\prime}}>,
m_{n},{\bf p}_{\eta^\prime}^{'}]}
{d{\bf p}_{\eta^\prime}^{'}}\right\}
$$
and
%formula(10)
\begin{equation}
  s=(E_{\pi^-}+E_t)^2-({\bf p}_{\pi^-}+{\bf p}_t)^2,
\end{equation}
%formula(11)
\begin{equation}
   E_t=M_A-\sqrt{(-{\bf p}_t)^2+(M_{A}-m_{N}+E)^{2}}.
\end{equation}
Here,
$d\sigma_{{\pi^-}p\to {\eta^\prime}n}[\sqrt{s},<m^*_{{\eta^\prime}}>,m_{n},{\bf p}_{\eta^\prime}^{'}]
/d{\bf p}_{\eta^\prime}^{'}$
is the off-shell inclusive differential cross section for the production of ${\eta^\prime}$ meson and neutron
with reduced mass $<m^*_{{\eta^\prime}}>$ and free mass $m_{n}$, respectively, and $\eta^\prime$ meson with in-medium momentum ${\bf p}_{{\eta^\prime}}^{'}$ in reaction (1) at the ${\pi^-}p$ center-of-mass energy $\sqrt{s}$;
$E_{\pi^-}$ and ${\bf p}_{\pi^-}$ are the total energy and momentum of the incident pion
($E_{\pi^-}=\sqrt{m^2_{\pi}+{\bf p}^2_{\pi^-}}$, $m_{\pi}$ is the free space pion mass);
$\rho({\bf r})$ and $P_A({\bf p}_t,E)$ are the local nucleon density and the
spectral function of the target nucleus A normalized to unity;
${\bf p}_t$ and $E$ are the internal momentum and removal energy of the struck target proton
involved in the collision process (1); $\sigma_{{\pi^-}N}^{\rm tot}$ is the total cross section of the
free ${\pi^-}N$ interaction
\footnote{$^)$We use in the subsequent calculations the value of $\sigma_{{\pi^-}N}^{\rm tot}=35$ mb
for the incident pion momentum of 1.7 GeV/c [13].}$^)$
;
$Z$ and $A$ are the numbers of protons and nucleons in
the target nucleus, and $M_{A}$  and $R$ are its mass and radius; and $\theta_{\eta^\prime}$ is the polar angle of
vacuum momentum ${\bf p}_{{\eta^\prime}}$ in the laboratory system with z-axis directed along the momentum
${\bf p}_{{\pi^-}}$ of the initial pion. Since the momenta of the outgoing neutrons in reaction (1)
are substantially greater than the typical average Fermi
momentum of $\sim$ 250 MeV/c of the target nucleus,
we neglect the correction of Eq.~(9) for
the Pauli blocking, leading to suppression of the available phase space.

        For the nuclear density $\rho({\bf r})$ in the cases of the $^{12}$C and $^{184}$W
target nuclei considered, we have employed, respectively, the harmonic oscillator
and the Woods-Saxon distributions [13]:
%FORMULA (12)
\begin{equation}
\rho({\bf r})={\rho}_{N}({\bf r})/A=\frac{(b/\pi)^{3/2}}{A/4}\left\{1+
\left[\frac{A-4}{6}\right]br^{2}\right\}\exp{(-br^2)},
\end{equation}
%formula(13)
\begin{equation}
 \rho({\bf r})=\rho_{0}\left[1+
\exp{\left(\frac{r-R_{1/2}}{a}\right)}\right]^{-1},
\end{equation}
where $b=0.355~{\rm fm}^{-2}$, $R_{1/2}=6.661~{\rm fm}$ and $a=0.55~{\rm fm}$.
For the $\eta^\prime$ production calculations in the case of the $^{12}$C target nucleus
the nuclear spectral function $P_A({\bf p}_t,E)$ was taken from Ref.~[18].
For the $^{184}$W target nucleus its single-particle part was assumed to be the same
as that for $^{208}$Pb [19]. This latter was taken from Ref.~[20]. The correlated part of the
spectral function for $^{184}$W was taken from Ref.~[18].

   Following Ref. [14], we assume that the off-shell inclusive differential cross section\\
$d\sigma_{{\pi^-}p\to {\eta^\prime}n}[\sqrt{s},<m^*_{{\eta^\prime}}>,m_{n},{\bf p}_{\eta^\prime}^{'}]
/d{\bf p}_{\eta^\prime}^{'}$
for $\eta^\prime$ creation in reaction (1) is equivalent to the respective on-shell cross section calculated for
the off-shell kinematics of this reaction as well as for the final $\eta^\prime$ and neutron in-medium mass
$<m^*_{{\eta^\prime}}>$ and free mass $m_{n}$, respectively.
    Taking into account Eq.~(16) from Ref. [14], we get the following expression for the
elementary in-medium differential cross section
$d\sigma_{{\pi^-}p\to {\eta^\prime}n}[\sqrt{s},<m^*_{{\eta^\prime}}>,m_{n},{\bf p}_{\eta^\prime}^{'}]
/d{\bf p}_{\eta^\prime}^{'}$:
%FORMULA (14)
\begin{equation}
\frac{d\sigma_{{\pi^{-}}p \to {\eta^\prime}n}[\sqrt{s},<m^*_{\eta^\prime}>,m_{n},
{\bf p}^{\prime}_{\eta^\prime}]}{d{\bf p}^{\prime}_{\eta^\prime}}=
\frac{\pi}{I_2[s,<m^*_{\eta^\prime}>,m_{n}]E^{\prime}_{\eta^\prime}}
\end{equation}
$$
\times
\frac{d\sigma_{{\pi^{-}}p \to {\eta^\prime}n}(\sqrt{s},<m^*_{\eta^\prime}>,m_{n},\theta^*_{\eta^\prime})}
{d{\bf \Omega}^*_{\eta^\prime}}
$$
$$
\times
\frac{1}{(\omega+E_t)}\delta\left[\omega+E_t-\sqrt{m_{n}^2+({\bf Q}+{\bf p}_t)^2}\right],
$$
where
%FORMULA (15)
\begin{equation}
I_2[s,<m^*_{\eta^\prime}>,m_{n}]=\frac{\pi}{2}
\frac{\lambda[s,(<m^*_{\eta^\prime}>)^{2},m_{n}^{2}]}{s},
\end{equation}
%FORMULA (16)
\begin{equation}
\lambda(x,y,z)=\sqrt{{\left[x-({\sqrt{y}}+{\sqrt{z}})^2\right]}{\left[x-
({\sqrt{y}}-{\sqrt{z}})^2\right]}},
\end{equation}
%FORMULA (17)
\begin{equation}
\omega=E_{\pi^-}-E^{\prime}_{\eta^\prime}, \,\,\,\,{\bf Q}={\bf p}_{\pi^-}-{\bf p}^{\prime}_{\eta^\prime}.
\end{equation}
Here,
$d\sigma_{{\pi^{-}}p \to {\eta^\prime}n}(\sqrt{s},<m^*_{\eta^\prime}>,m_{n},\theta^*_{\eta^\prime})
/d{\bf \Omega}^*_{\eta^\prime}$
is the off-shell differential cross section for the production of $\eta^\prime$ mesons with mass $<m^*_{\eta^\prime}>$
in reaction (1) under the polar angle $\theta^*_{\eta^\prime}$ in the ${\pi^-}p$ c.m.s., which is assumed to be
isotropic in our calculations of $\eta^\prime$ meson creation in ${\pi^-}A$ collisions from this reaction:
%FORMULA (18)
\begin{equation}
\frac{d\sigma_{{\pi^{-}}p \to {\eta^\prime}n}(\sqrt{s},<m^*_{\eta^\prime}>,m_{n},\theta^*_{\eta^\prime})}
{d{\bf \Omega}^*_{\eta^\prime}}=\frac{\sigma_{{\pi^{-}}p \to {\eta^\prime}n}(\sqrt{s},\sqrt{s^*_{\rm th}})}{4\pi}.
\end{equation}
Here, $\sigma_{{\pi^{-}}p \to {\eta^\prime}n}(\sqrt{s},\sqrt{s^*_{\rm th}})$ is the ``in-medium" total cross section
of reaction (1) having the threshold energy $\sqrt{s^*_{\rm th}}=<m^*_{\eta^\prime}>+m_{n}$.
In line with the above, it is equivalent to the vacuum cross section
$\sigma_{{\pi^{-}}p \to {\eta^\prime}n}(\sqrt{s},\sqrt{s_{\rm th}})$, in which the free threshold energy
$\sqrt{s_{\rm th}}=m_{\eta^\prime}+m_{n}=1.897$ GeV is replaced by the in-medium threshold
energy $\sqrt{s^*_{\rm th}}$. For the free total cross section
$\sigma_{{\pi^{-}}p \to {\eta^\prime}n}(\sqrt{s},\sqrt{s_{\rm th}})$ we have used the following parametrization
suggested in Ref. [14]:
%formula(19)
\begin{equation}
\sigma_{{\pi}^-p \to {\eta^\prime}n}(\sqrt{s},\sqrt{s_{\rm th}})=\left\{
\begin{array}{ll}
	223.5\left(\sqrt{s}-\sqrt{s_{\rm th}}\right)^{0.352}~[{\rm {\mu}b}]
	&\mbox{for $0 < \sqrt{s}-\sqrt{s_{\rm th}} \le 0.354~{\rm GeV}$}, \\
	&\\
                   35.6/\left(\sqrt{s}-\sqrt{s_{\rm th}}\right)^{1.416}~[{\rm {\mu}b}]
	&\mbox{for $\sqrt{s}-\sqrt{s_{\rm th}} > 0.354~{\rm GeV}$}.
\end{array}
\right.	
\end{equation}
As shown in Fig. 1, the parametrization (19) (solid line) fits well the existing sets of experimental
data [21] for ${\pi^-}p \to {\eta^\prime}n$ (full triangles) and ${\pi^+}n \to {\eta^\prime}p$ (full circles)
reactions. Looking at this figure, one can see that the cross section $\sigma_{{\pi}^-p \to {\eta^\prime}n}$
amounts approximately to 110 $\mu$b for the initial pion momentum of 1.7 GeV/c.
%%%%%%%%%%%%%%%%%%%%%%%%%%%%%%%%%%%%%%%%%%%%%%%%%%%%%%%%%%%
\begin{figure}[htb]
\begin{center}
\includegraphics[width=12.0cm]{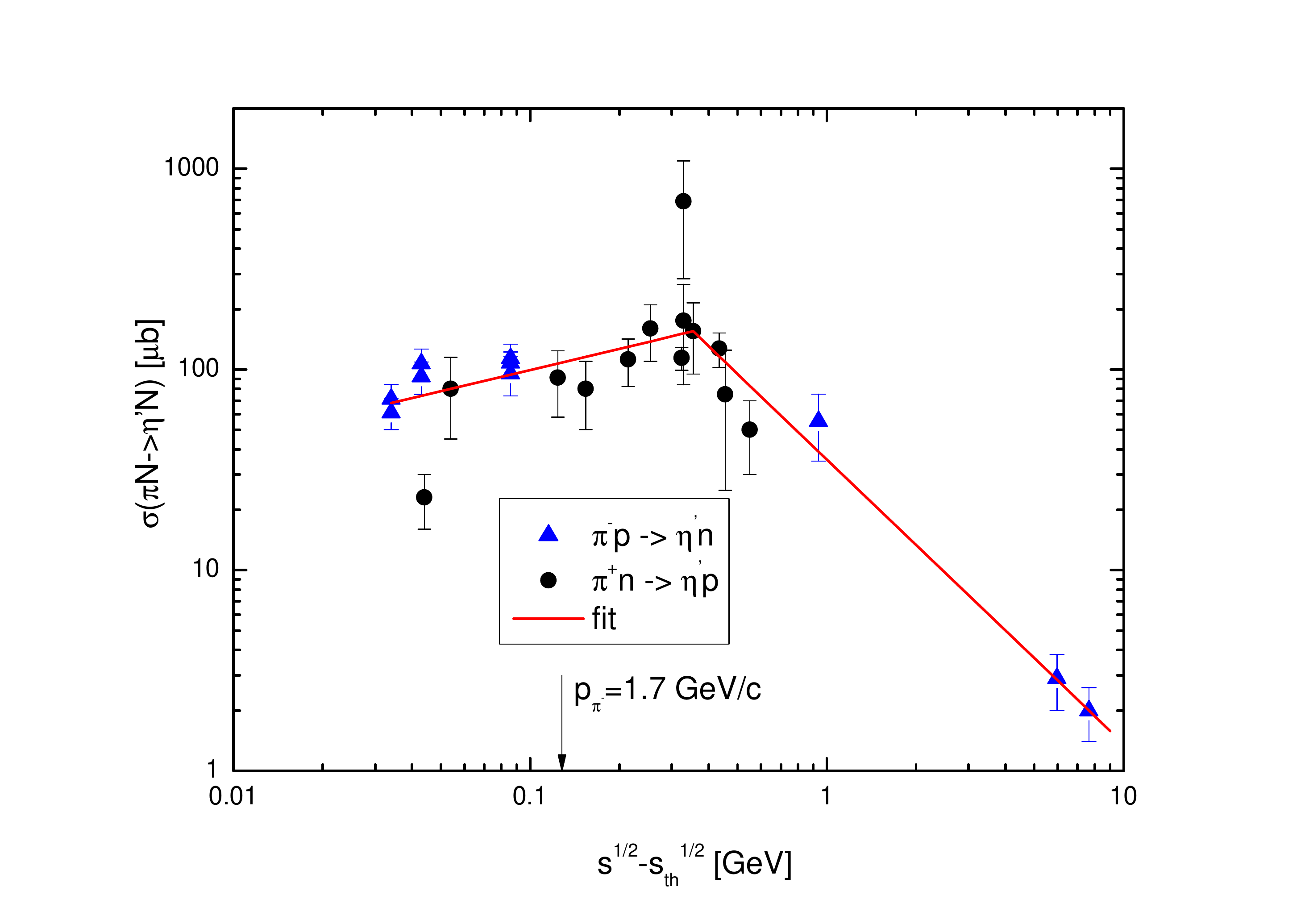}
\vspace*{-2mm} \caption{(color online) The total cross sections for the reactions
$\pi^-p \to {\eta^\prime}n$ and $\pi^+n \to {\eta^\prime}p$
as functions of the excess energy $\sqrt{s}-\sqrt{s_{\rm th}}$.
The arrow indicates the excess energy $\sqrt{s}-\sqrt{s_{\rm th}}=128$ MeV corresponding to
the incident pion momentum of 1.7 GeV/c and a free target nucleon at rest.}
\label{void}
\end{center}
\end{figure}
%%%%%%%%%%%%%%%%%%%%%%%%%%%%%%%%%%%%%%%%%%%%%%%%%%%%%%%%%%%%%%%%%%%%%%%%%%%%%%%%%%%%%%%%%%%%%
It is worth noting that for this momentum the cross section
$\sigma_{{\pi}^-p \to {\eta^\prime}n}$ is larger by about two orders of magnitude
than the total cross section
$\sigma_{{\gamma}N \to {\eta^\prime}N}$ for $\eta^\prime$ photoproduction off the nucleon
at incident photon energies around 2 GeV [3, 22, 23], studied in the CBELSA/TAPS
$\eta^\prime$ photoproduction experiments [3--5].
Therefore, the collected statistics in pion-induced $\eta^\prime$ meson production experiment
at beam momentum of 1.7 GeV/c is expected to be essentially higher than in the above experiments.
This offers the possibility to investigate the $\eta^\prime$ effective scalar potential depth at
substantially lower $\eta^\prime$ momenta ($\sim$ 200--300 MeV/c) compared to those, studied in
the experiments [3--5], via the measurements of the $\eta^\prime$ momentum distributions in
${\pi^-}A$ reactions at modern experimental facilities such as the GSI pion beam facility and J-PARC.

    In Eqs.~(6)--(8) we assume that the
direction of the three-momentum of the $\eta^\prime$ meson
remains unchanged as it propagates from its production point inside the nucleus in the rather
weak nuclear mean-field, considered in this paper, to the vacuum outside the nucleus.
In this case, the quantities
$\left<d\sigma_{{\pi^-}p\to {\eta^\prime}n}({\bf p}_{\pi^-},
{\bf p}^{\prime}_{\eta^\prime})/d{\bf p}^{\prime}_{\eta^\prime}\right>_A$
and
$d{\bf p}^{\prime}_{\eta^\prime}/d{\bf p}_{\eta^\prime}$,
entering into Eq. (5), can be accounted for in our calculations as
$\left<d\sigma_{{\pi^-}p\to {\eta^\prime}n}(p_{\pi^-},
p^{\prime}_{\eta^\prime},\theta_{\eta^\prime})/p^{\prime2}_{\eta^\prime}dp^{\prime}_{\eta^\prime}d{\bf \Omega}_{\eta^\prime}\right>_A$
and
$p^{\prime}_{\eta^\prime}/{p}_{\eta^\prime}$, where
${\bf \Omega}_{\eta^\prime}(\theta_{\eta^\prime},
\varphi_{\eta^\prime})={\bf p}_{\eta^\prime}/p_{\eta^\prime}$. Here, $\varphi_{\eta^\prime}$ is the
azimuthal angle of the $\eta^\prime$ vacuum momentum ${\bf p}_{\eta^\prime}$ in the lab system.
Similar to the $\phi$ meson production in $\pi^-A$ ($A=^{12}$C and $^{184}$W) collisions at incident
pion momentum of 1.7 GeV/c in the HADES experiment [9], we will consider the $\eta^\prime$ momentum
distribution on these target nuclei in the acceptance window
${\Delta}{\bf \Omega}_{\eta^\prime}$=$10^{\circ} \le \theta_{\eta^\prime} \le 45^{\circ}$ and
$0 \le \varphi_{\eta^\prime} \le 2{\pi}$.
Integrating the full inclusive differential cross section (5) over this range,
we can represent the differential cross section for $\eta^\prime$ meson
production in ${\pi^-}A$ collisions from the direct process (1), corresponding to the kinematical
conditions of the HADES $\phi$ meson production experiment [9], in the following form:
%formula(20)
\begin{equation}
\frac{d\sigma_{{\pi^-}A\to {\eta^\prime}X}^{({\rm prim})}
(p_{\pi^-},p_{\eta^\prime})}{dp_{\eta^\prime}}=
\int\limits_{{\Delta}{\bf \Omega}_{\eta^\prime}}^{}d{\bf \Omega}_{\eta^\prime}
\frac{d\sigma_{{\pi^-}A\to {\eta^\prime}X}^{({\rm prim})}
({\bf p}_{\pi^-},{\bf p}_{\eta^\prime})}{d{\bf p}_{\eta^\prime}}p_{\eta^\prime}^2
\end{equation}
$$
=2{\pi}\left(\frac{Z}{A}\right)\left(\frac{p_{\eta^\prime}}{p^{\prime}_{\eta^\prime}}\right)
\int\limits_{\cos45^{\circ}}^{\cos10^{\circ}}d\cos{{\theta_{\eta^\prime}}}I_{V}[A,\theta_{\eta^\prime}]
\left<\frac{d\sigma_{{\pi^-}p\to {\eta^\prime}n}(p_{\pi^-},
p^{\prime}_{\eta^\prime},\theta_{\eta^\prime})}{dp^{\prime}_{\eta^\prime}d{\bf \Omega}_{\eta^\prime}}\right>_A.
$$
To get a better impression of the size of the effect of $\eta^\prime$ meson in medium mass shift on its
yield in ${\pi^-}C \to {\eta^\prime}X$ and ${\pi^-}W \to {\eta^\prime}X$ reactions as well as of the strength
of this yield in another laboratory $\eta^\prime$ emission polar angular ranges, we will also calculate the
$\eta^\prime$ momentum distribution in these reactions with $\eta^\prime$ going into the laboratory solid
angle
${\Delta}{\bf \Omega}_{\eta^\prime}$=$0^{\circ} \le \theta_{\eta^\prime} \le 10^{\circ}$ and
$0 \le \varphi_{\eta^\prime} \le 2{\pi}$.

%%%%%%%%%%%%%%%%%%%%%%%%%%%%%%%%%%%%%%%%%%%%%%%%%%%%%%%%%%%
\begin{figure}[!h]
\begin{center}
\includegraphics[width=16.0cm]{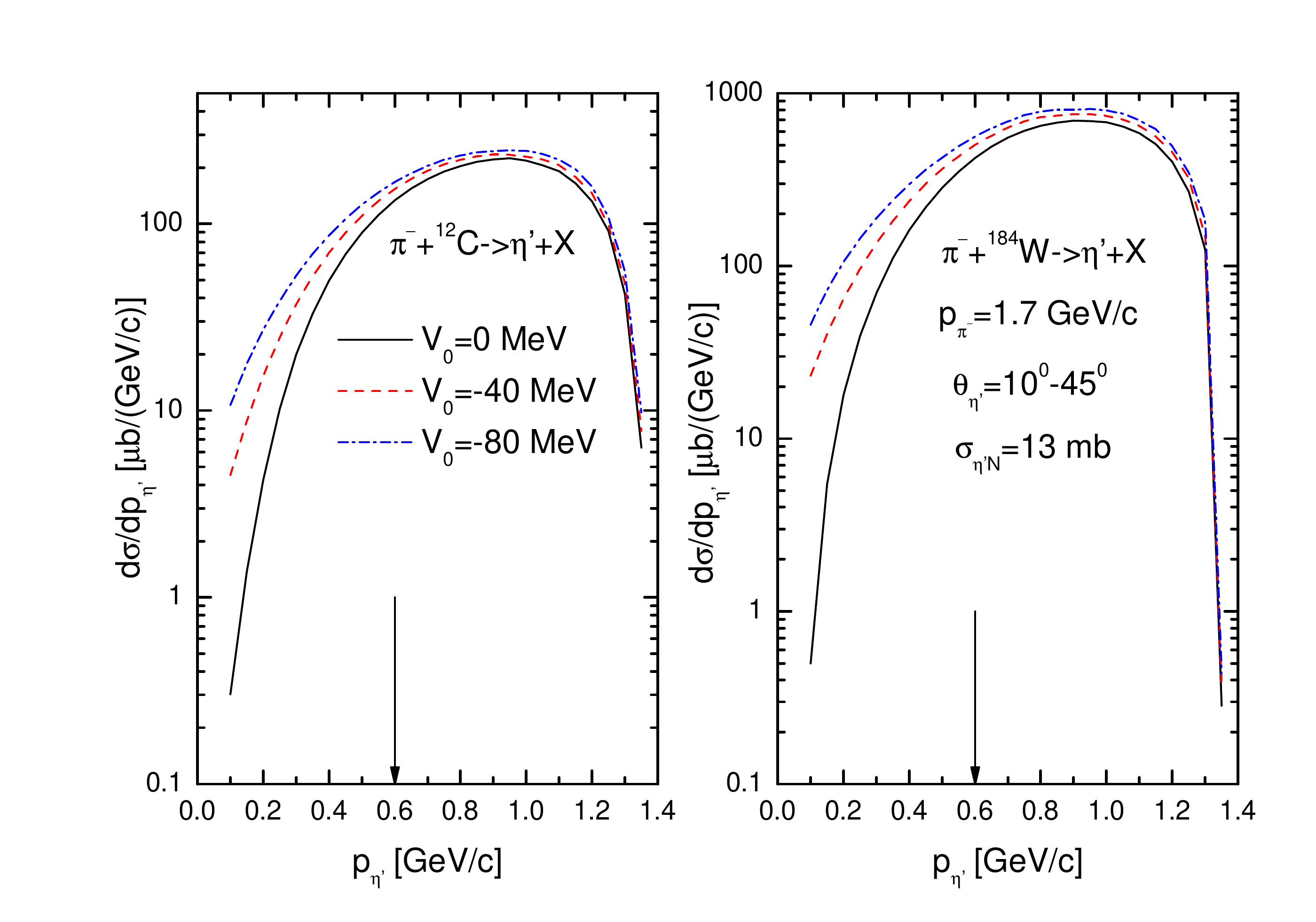}
\vspace*{-2mm} \caption{(color online) Momentum differential cross sections for the production of $\eta^\prime$
mesons from the primary ${\pi^-}p \to {\eta^\prime}n$ channel in the laboratory polar angular range of
10$^{\circ}$--45$^{\circ}$ in the interaction of $\pi^-$ mesons of momentum of 1.7 GeV/c with $^{12}$C
(left panel) and $^{184}$W (right panel) nuclei for different values of the
$\eta^\prime$ meson effective scalar potential $V_0$ at density $\rho_0$ indicated in the inset
in the scenario with $\sigma_{{\eta^\prime}N}=13$ mb.
The arrows indicate the boundary between the low-momentum and high-momentum
regions of 0.1--0.6 GeV/c and 0.6--1.35 GeV/c, respectively, of the $\eta^\prime$ spectra.}
\label{void}
\end{center}
\end{figure}
%%%%%%%%%%%%%%%%%%%%%%%%%%%%%%%%%%%%%%%%%%%%%%%%%%%%%%%%%%%%%%%%%%%%%%%%%%%%%%%%%%%%%%%%%%%%%
\begin{figure}[!h]
\begin{center}
\includegraphics[width=16.0cm]{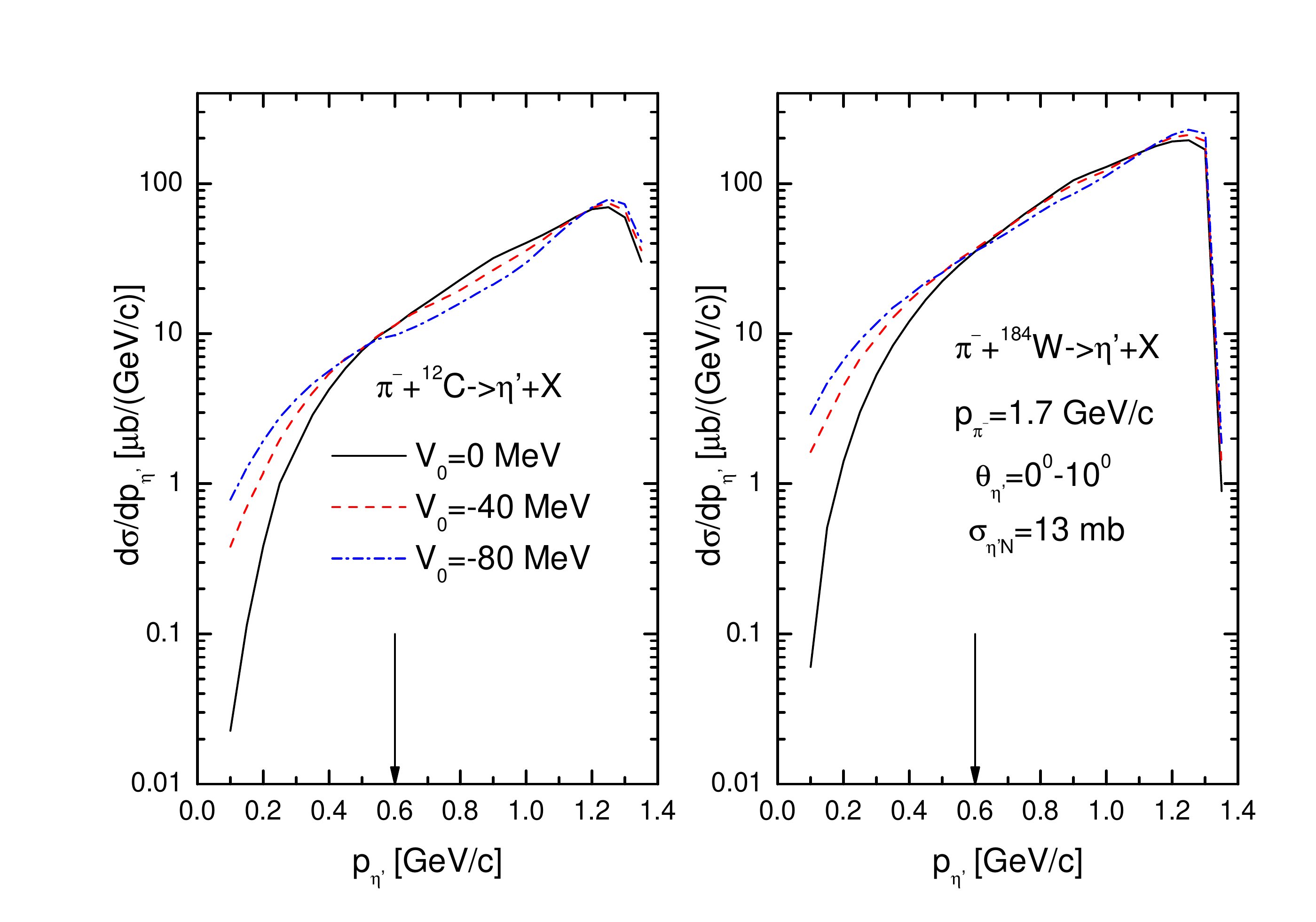}
\vspace*{-2mm} \caption{(color online) Momentum differential cross sections for the production of $\eta^\prime$
mesons from the primary ${\pi^-}p \to {\eta^\prime}n$ channel in the laboratory polar angular range of
0$^{\circ}$--10$^{\circ}$ in the interaction of $\pi^-$ mesons of momentum of 1.7 GeV/c with $^{12}$C
(left panel) and $^{184}$W (right panel) nuclei for different values of the
$\eta^\prime$ meson effective scalar potential $V_0$ at density $\rho_0$ indicated in the inset
in the scenario with $\sigma_{{\eta^\prime}N}=13$ mb.
The arrows indicate the boundary between the low-momentum and high-momentum
regions of 0.1--0.6 GeV/c and 0.6--1.35 GeV/c, respectively, of the $\eta^\prime$ spectra.}
\label{void}
\end{center}
\end{figure}
%%%%%%%%%%%%%%%%%%%%%%%%%%%%%%%%%%%%%%%%%%%%%%%%%%%%%%%%%%%
\begin{figure}[!h]
\begin{center}
\includegraphics[width=18.0cm]{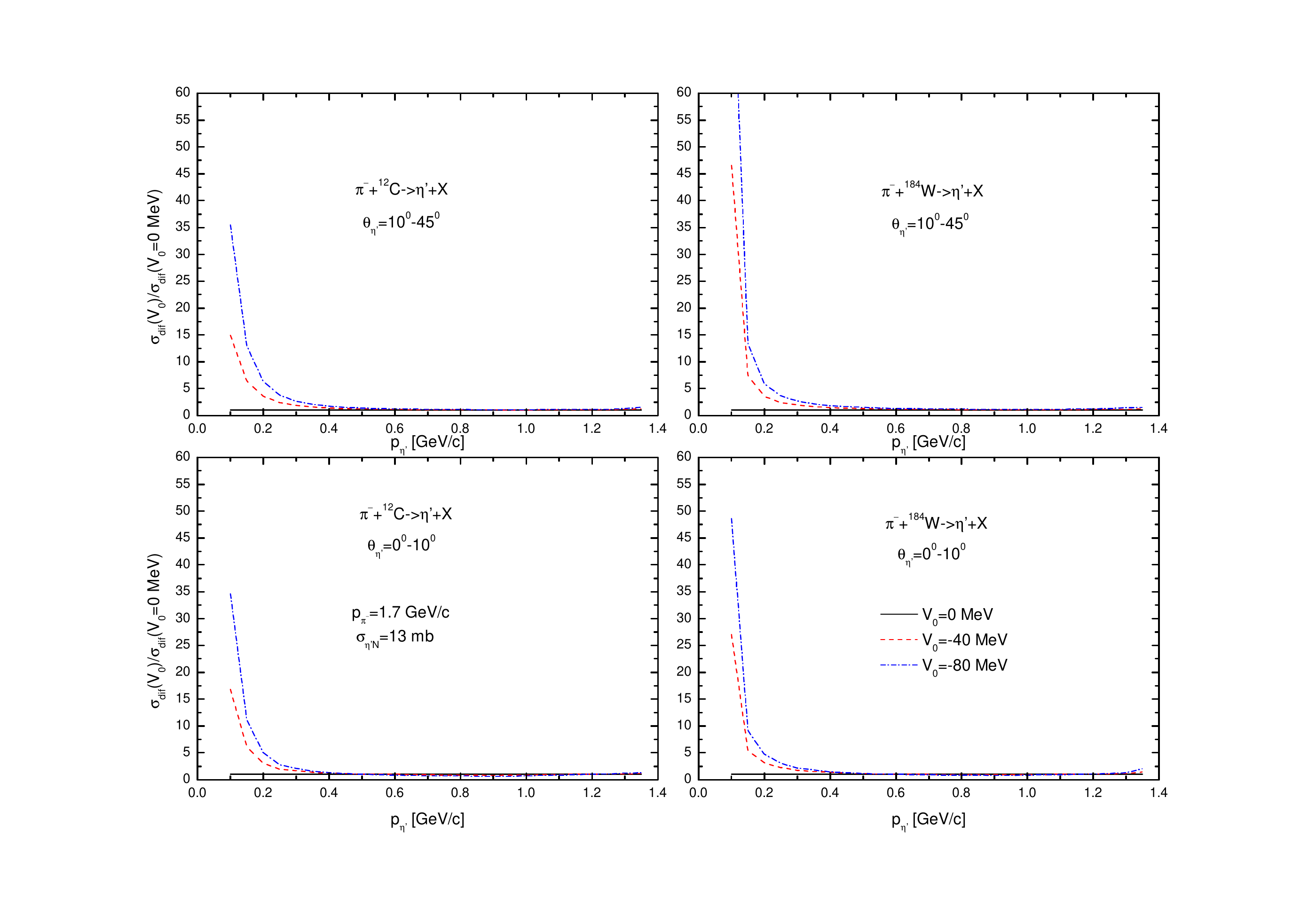}
\vspace*{-2mm} \caption{(color online) Ratio between the differential cross sections of $\eta^\prime$
production on C and W target nuclei as a function of the $\eta^\prime$ momentum
at $\eta^\prime$ laboratory angles 10$^{\circ}$--45$^{\circ}$
(upper two panels) and 0$^{\circ}$--10$^{\circ}$ (lower two panels) by 1.7 GeV/c $\pi^-$ mesons in
primary $\pi^-p \to {\eta^\prime}n$ reactions proceeding on off-shell target protons, calculated
with and without the $\eta^\prime$ in-medium mass shift, with values shown in the inset,
in the scenario with $\sigma_{{\eta^\prime}N}=13$ mb.}
\label{void}
\end{center}
\end{figure}
%%%%%%%%%%%%%%%%%%%%%%%%%%%%%%%%%%%%%%%%%%%%%%%%%%%%%%%%%%%%%%%%%%%%%%%%%%%%%%%%%%%%%%%%%%%%%
%%%%%%%%%%%%%%%%%%%%%%%%%%%%%%%%%%%%%%%%%%%%%%%%%%%%%%%%%%%
\begin{figure}[!h]
\begin{center}
\includegraphics[width=18.0cm]{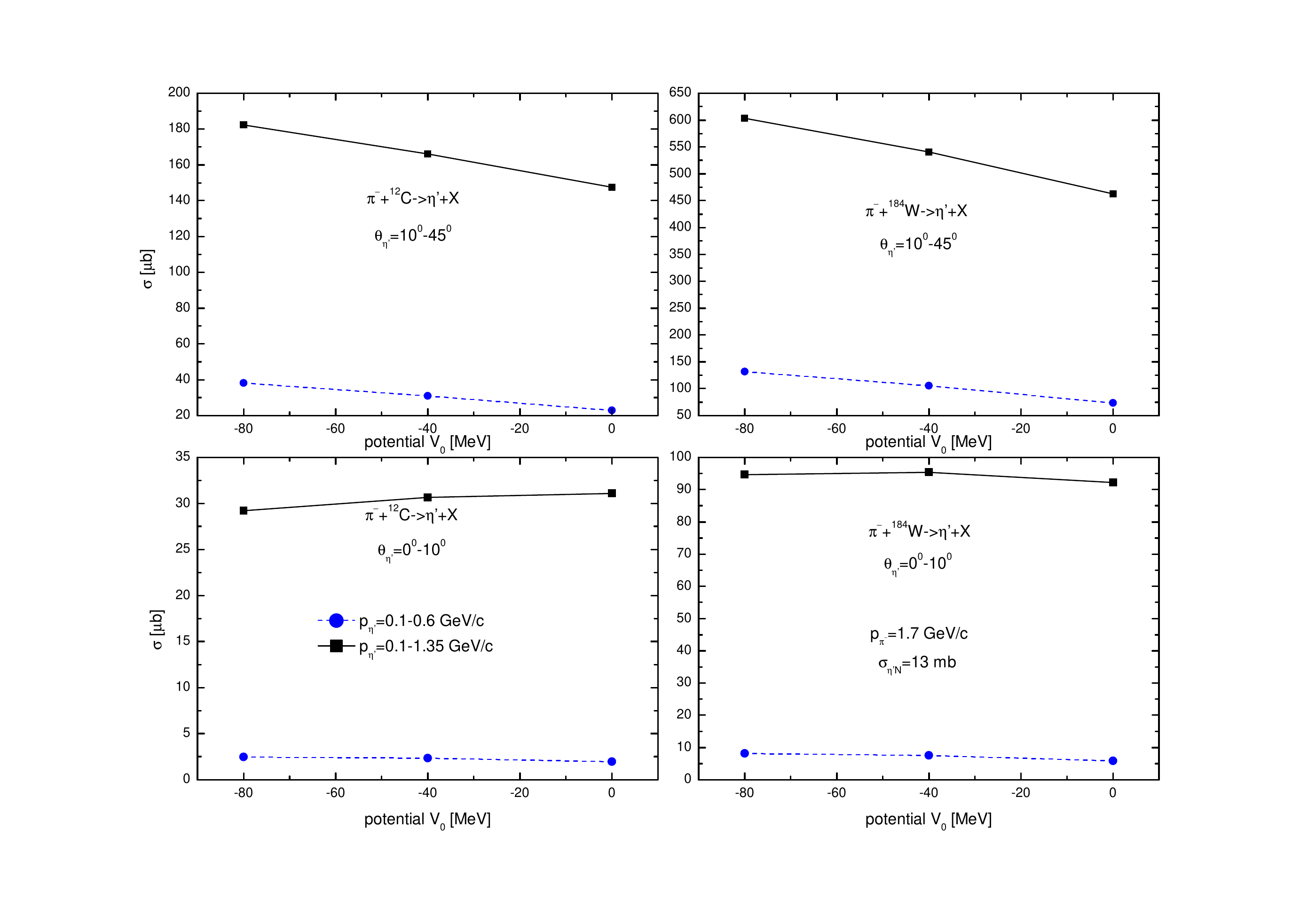}
\vspace*{-2mm} \caption{(color online) The total cross sections for the production of $\eta^\prime$
mesons from the primary ${\pi^-}p \to {\eta^\prime}n$ channel on C and W target nuclei with
momenta of 0.1--0.6 GeV/c and 0.1--1.35 GeV/c
in the laboratory polar angular ranges of 10$^{\circ}$--45$^{\circ}$ (upper two panels) and
0$^{\circ}$--10$^{\circ}$ (lower two panels) by 1.7 GeV/c $\pi^-$ mesons
in the scenario with $\sigma_{{\eta^\prime}N}=13$ mb
as functions of the effective scalar $\eta^\prime$ potential $V_0$ at normal nuclear density. The lines
are to guide the eye.}
\label{void}
\end{center}
\end{figure}
%%%%%%%%%%%%%%%%%%%%%%%%%%%%%%%%%%%%%%%%%%%%%%%%%%%%%%%%%%%%%%%%%%%%%%%%%%%%%%%%%%%%%%%%%%%%%

\section*{3 Results and discussion}

\hspace{0.5cm} First of all, we consider the momentum dependence of the absolute
$\eta^\prime$ meson yield from the direct process (1). Figures 2 and 3 give momentum spectra
of $\eta^\prime$ mesons produced in the interaction of 1.7 GeV/c $\pi^-$ mesons with
$^{12}$C and  $^{184}$W nuclei at laboratory angles of 10$^{\circ}$--45$^{\circ}$ and
0$^{\circ}$--10$^{\circ}$, respectively. In these figures, the curves -- the respective momentum
differential cross sections -- represent the results of calculations on the basis of expression (20)
for three values of the $\eta^\prime$ meson shift at a normal nuclear density: 0, -40, and -80 MeV
in the scenario of collisional broadening of the $\eta^\prime$ meson characterized by the value of
$\sigma_{{\eta^\prime}N}=13$ mb. The absolute values of the differential cross sections show
a rather wide variation for the mass shift range of $V_0=0$ to -80 MeV in the
low-momentum region of 0.1--0.6 GeV/c and especially at $\eta^\prime$ momenta $\sim$ 0.2--0.3 GeV/c
(where they have yet a measurable strength $\sim$ 1 $\mu$b/(GeV/c) and larger) for both target nuclei
and for both laboratory polar $\eta^\prime$ production angular ranges of 0$^{\circ}$--10$^{\circ}$
and 10$^{\circ}$--45$^{\circ}$.

 More detailed information about the sensitivity of the cross sections to the $\eta^\prime$ mass shift
is contained in Fig. 4, where we show the momentum dependence of the ratio
\footnote{$^)$We remind that a comparison of a similar differential cross section ratio calculated
at various values of the $\eta^\prime$ in-medium mass shift with data on $\eta^\prime$ photoproduction
off C and Nb nuclei was used in [3, 4] to extract this shift at a normal nuclear density.}$^)$
between the differential cross sections for $\eta^\prime$ production on $^{12}$C and  $^{184}$W
target nuclei, calculated for different values for $\eta^\prime$ mass shift at saturation density
and presented in Figs. 2, 3, and the respective differential cross section, determined without this shift.
One can see that at low ($\sim$ 0.2--0.3 GeV/c) $\eta^\prime$ momenta there are sizeable and experimentally
distinguishable differences between the options $V_0=0$ MeV, $V_0=-40$ MeV, and $V_0=-80$ MeV for
$\eta^\prime$ in-medium mass shift for all considered target nuclei and laboratory
polar $\eta^\prime$ production angular domains. Thus, the $\eta^\prime$ momentum distributions are enhanced
at mass shift $V_0=-40$ MeV by about a factor of 2.4 as compared to those obtained without this shift at momentum
of 0.25 GeV/c.  When going from $V_0=-40$ MeV to $V_0=-80$ MeV, the corresponding enhancement factor is of about
1.5 at this $\eta^\prime$ momentum. Accounting for the above enhancement factors, one may hope that
even smaller mass
shifts ($V_0 \sim -20$ MeV) will probably be also experimentally accessible via the measurements of low-momentum
parts of $\eta^\prime$ spectra in near-threshold ${\pi^-}A$ interactions.

  As also seen from Figs. 2 and 3, the $\eta^\prime$ production cross section at laboratory angles of
$\theta_{{\eta^\prime}}=$10$^{\circ}$--45$^{\circ}$ and $\eta^\prime$ momenta $\sim$ 0.2--0.3 GeV/c
is greater than that at $\eta^\prime$ emission angles of $\theta_{{\eta^\prime}}=$0$^{\circ}$--10$^{\circ}$
for these momenta by about of one order of magnitude. This makes the measurement of $\eta^\prime$
with momenta $\sim$ 0.2--0.3 GeV/c in ${\pi^-}A$ reactions for initial momentum of 1.7 GeV/c,
for example, at GSI pion beam facility, using the HADES spectrometer
\footnote{$^)$In the CBELSA/TAPS $\eta^\prime$ photoproduction experiments [3--6] $\eta^\prime$ mesons
were identified via the
$\eta^\prime \to {\pi^0}{\pi^0}{\eta} \to {\pi^0}{\pi^0}{\gamma}{\gamma} \to 6{\gamma}$ decay chains with
a branching ratio of 8.5 \%. The measurement of $\eta^\prime$ mesons at HADES has only recently become
possible. So far, HADES spectrometer could only detect charged particles, but recently an electromagnetic
calorimeter was added to allow for the detection of photons. Thus, detection of $\eta^\prime$ mesons
has become possible here [24]. At HADES one would probably identify the $\eta^\prime$ meson via the
decay $\eta^\prime \to {\pi^+}{\pi^-}{\eta} \to {\pi^+}{\pi^-}{\gamma}{\gamma}$ with a branching
ratio of 16.6 \%. An alternative decay mode which may be easier experimentally because of the lower
background could be
$\eta^\prime \to {\pi^+}{\pi^-}{\eta} \to {\pi^+}{\pi^-}{\pi^+}{\pi^-}{\pi^0} \to 2{\pi^+}2{\pi^-}2{\gamma}$
with a branching ratio of 11.9 \% [24].}$^)$
,
with aim of distinguishing at these momenta at least between zero,
weak ($V_0 \sim -40$ MeV) and relatively weak ($V_0 \sim -80$ MeV) $\eta^\prime$ mass shifts
(or $\eta^\prime$ effective scalar potentials) in cold nuclear matter quite promising.
It is interesting to note
that the model differential cross sections for $\eta^\prime$ photoproduction off $^{12}$C nucleus, adopted in
Ref. [3] for extracting the $\eta^\prime$ scalar potential $V_0$, are less than those presented in
Fig. 2 (left) by about two orders of magnitude at momenta $\sim$ 0.2--0.3 GeV/c. This implies that the differential
cross sections for $\eta^\prime$ production in ${\pi^-}A$ collisions can be measured at these momenta with a substantially higher accuracy than that reached in the experiments [3, 4]. This would definitely permit
evaluating the $\eta^\prime$ scalar potential $V_0$ at $\eta^\prime$ momenta $\sim$ 0.2--0.3 GeV/c in these collisions.

      The sensitivity of the strength of the low-momentum parts of the $\eta^\prime$ spectra to the scalar
potential $V_0$, shown in Figs. 2, 3, can be exploited to infer the correlation between the above strength
and this potential from such integral measurements as the measurements of the
total cross sections for $\eta^\prime$ production in ${\pi^-}^{12}$C and  ${\pi^-}^{184}$W collisions
by 1.7 GeV/c pions in the low-momentum (0.1--0.6 GeV/c) and full-momentum (0.1--1.35 GeV/c) regions.
These cross sections, calculated for the $\eta^\prime$ laboratory polar angular domains of
0$^{\circ}$--10$^{\circ}$ and 10$^{\circ}$--45$^{\circ}$
by integrating Eq. (20) over the $\eta^\prime$ momentum $p_{\eta^\prime}$
in these regions in the scenario with $\sigma_{{\eta^\prime}N}=13$ mb, as functions of the potential $V_0$
are shown in Fig. 5. It is seen from this figure that due to larger cross sections and higher sensitivity
to this potential the $\eta^\prime$ polar angular domain of 10$^{\circ}$--45$^{\circ}$
is more favorable than the forward $\eta^\prime$ production polar angular range of
0$^{\circ}$--10$^{\circ}$. In this angular domain the highest sensitivity of the $\eta^\prime$
production cross section to the scalar potential $V_0$ is observed, as was expected, in the low-momentum
region of 0.1--0.6 GeV/c. Thus, for example, the $\eta^\prime$ total cross section is enhanced here at
$V_0=-80$ MeV by a factor of about 1.8 compared to that obtained in the scenario with zero $\eta^\prime$
potential. Whereas, in the full $\eta^\prime$ momentum range of 0.1--1.35 GeV/c the respective enhancement
factor is about 1.3. Therefore, a comparison of the above "integral" results with the experimentally determined
total $\eta^\prime$ creation cross sections in the low-momentum region of 0.1--0.6 GeV/c will also allow one
to get the definite information about the average effective scalar potential $V_0$ (or about the $\eta^\prime$
in-medium mass shift) at normal nuclear matter density $\rho_0$ in this region.

   Accounting for the above consideration, we come to the conclusion that the near-threshold $\eta^\prime$
momentum distribution and $\eta^\prime$ total cross section measurements at momenta $\sim$ 0.2--0.3 GeV/c
and 0.1--0.6 GeV/c, respectively, in ${\pi^-}A$ reactions might permit to shed light on the possible
$\eta^\prime$ in-medium mass shift (or on the scalar $\eta^\prime$--nucleus optical potential) at these
momenta. Such measurements could be performed in the future at the GSI pion beam facility or at
J-PARC
\footnote{$^)$It should be pointed out that the ($\pi^+, p$) and ($\pi^-, n$) reactions were proposed earlier [25]
for the experimental search of $\eta^\prime$-nuclear bound states at J-PARC.}$^)$
.

\section*{4 Conclusions}

\hspace{0.5cm} Our present study was aimed at studying the possibility of extracting information about
the in-medium change of the $\eta^\prime$ meson mass (or about the effective $\eta^\prime$
scalar nuclear potential) at low meson momenta. For this, we have calculated the differential and total cross
sections for $\eta^\prime$ production off carbon and tungsten nuclei at laboratory angles of
0$^{\circ}$--10$^{\circ}$ and 10$^{\circ}$--45$^{\circ}$ by $\pi^-$ mesons with momentum of 1.7 GeV/c,
which is close to the threshold momentum for $\eta^\prime$ meson production off the free target proton at rest.
We have performed these calculations on the basis of the first-collision model, which describes incoherent
$\eta^\prime$ meson creation in single collisions of incident $\pi^-$ mesons with intranuclear target protons
and takes into account three different options for its in-medium mass shift at saturation density $\rho_0$.
The intrinsic properties of $^{12}$C and  $^{184}$W target nuclei have been described in terms of their spectral functions, which account for the momenta of target protons and the energies of their separation from the nuclei.
An inspection of the results of these calculations has shown that the present $\eta^\prime$ production cross sections are larger than those, studied in Ref. [3] in the $\eta^\prime$ photoproduction reactions, by about two orders of magnitude. At the same time, the calculations have shown that the high absolute values of the differential
and total $\eta^\prime$ pion-induced production cross sections in the
momentum ranges of 0.2--0.3 GeV/c and 0.1--0.6 GeV/c, respectively, possess a high sensitivity
to changes in the in-medium shift of the $\eta^\prime$ mass. This offers the possibility of evaluating
the above shift at these momenta from the respective data,
which could be taken in future experiments at the GSI pion beam facility or at J-PARC.
\\
\\
{\bf Acknowledgments}
\\
\\
The author gratefully acknowledges Volker Metag for careful reading of the manuscript and valuable
comments on it, improving its quality.
\\
\\

%%%%%%%%%%%%%%%%%%%%%%%%%%%%%%%%%%%%%%%%%%%%%%%%%%%%%%%%%%%%%%%%
\end{document}